\pdfoutput=1
\documentclass[aps, prl, floatfix, twocolumn, amsmath, superscriptaddress, showpacs, 10pt]{revtex4-1}

\usepackage{graphicx}
\usepackage{color}

\DeclareMathOperator{\arctanh}{arctanh}

\newcommand{\PRLsec}[1]{\noindent\textit{#1}---}

\begin{document}
\title{A Continuum Generalization of the Ising Model}
\author{Haley A. Yaple}
\email[email: ]{haleyyaple@u.northwestern.edu}
\affiliation{Department of Engineering Sciences and Applied Mathematics, Northwestern University, Evanston, IL 60208, USA}

\author{Daniel M. Abrams}
\affiliation{Department of Engineering Sciences and Applied Mathematics, Northwestern University, Evanston, IL 60208, USA}
\affiliation{Northwestern Institute on Complex Systems, Northwestern University, Evanston, IL 60208, USA}
\date{\today}

\begin{abstract}
The Lenz-Ising model has served for almost a century as a basis for understanding ferromagnetism, and has become a paradigmatic model for phase transitions in statistical mechanics.  While retaining the Ising energy arguments, we use techniques previously applied to sociophysics to propose a continuum model. Our formulation results in an integro-differential equation that has several advantages over the traditional version: it allows for asymptotic analysis of phase transitions, material properties, and the dynamics of the
formation of magnetic domains. 
\end{abstract}

\maketitle

\PRLsec{Introduction}%
The Ising Model is a widely-studied model for magnetic phenomena \cite{domb1974,dejongh1974} which posits that each particle in a material has associated with it a binary magnetic polarity, or ``spin,'' that may flip to reduce the energy of the system. Derived as an equilibrium model in statistical mechanics, results for one-dimensional nearest-neighbor coupling were presented by Ising in his 1924 doctoral thesis \cite{ising1925}. An exact solution for two-dimensional nearest-neighbor coupling was found in 1944 by Onsager \cite{onsager1944}, spurring a flurry of research in the mid-twentieth century. Since then the Ising model has motivated thousands of peer-reviewed publications and become fundamental to understanding phase transitions \cite{stanley1999}.

While research on the Ising model has helped to provide insight into ferromagnetism and more general phase transitions, the statistical-mechanical approach has several disadvantages:  traditional solutions to the Ising model cannot give information about the dynamics of phase transitions and approach to equilibrium;  most results other than mean-field require laborious series expansions; no closed-form solution has been found in three dimensions, and such solutions are unlikely to be possible for irregular lattices.

Some of these drawbacks are remedied by simulating the Ising model numerically, following Monte-Carlo algorithms \cite{metropolis1953,binder1976}. However, numerical approaches do not provide much insight into dynamics, and the time required for simulation can be extremely long for large systems near critical points.

In this paper, we construct a continuous, deterministic model for ferromagnetism based on the energy arguments of the Ising model. This allows us to employ the tools of dynamical systems and perturbation theory to investigate time-dependence and asymptotic behavior for magnetization phenomena, as well as to recover mean-field results using new methods.

\PRLsec{Our Model}%
In the statistical mechanics formulation of the Ising model, the system energy is given by
\begin{align}
  E = -\frac{1}{2}\sum_i\sum_j J_{ij}s_is_j-\mu H\sum_i s_i~,
  \label{totenergy}
\end{align}
where $J_{ij}$ defines the magnitude of coupling between spins $i$ and $j$ (their interaction energy), $H$ is the external magnetic field, and $\mu$ is the magnetic moment, which we set to 1 for convenience. 
We will take two continuum limits of Eq.~\eqref{totenergy}. 
\begin{figure}
\includegraphics[width=.9\linewidth]{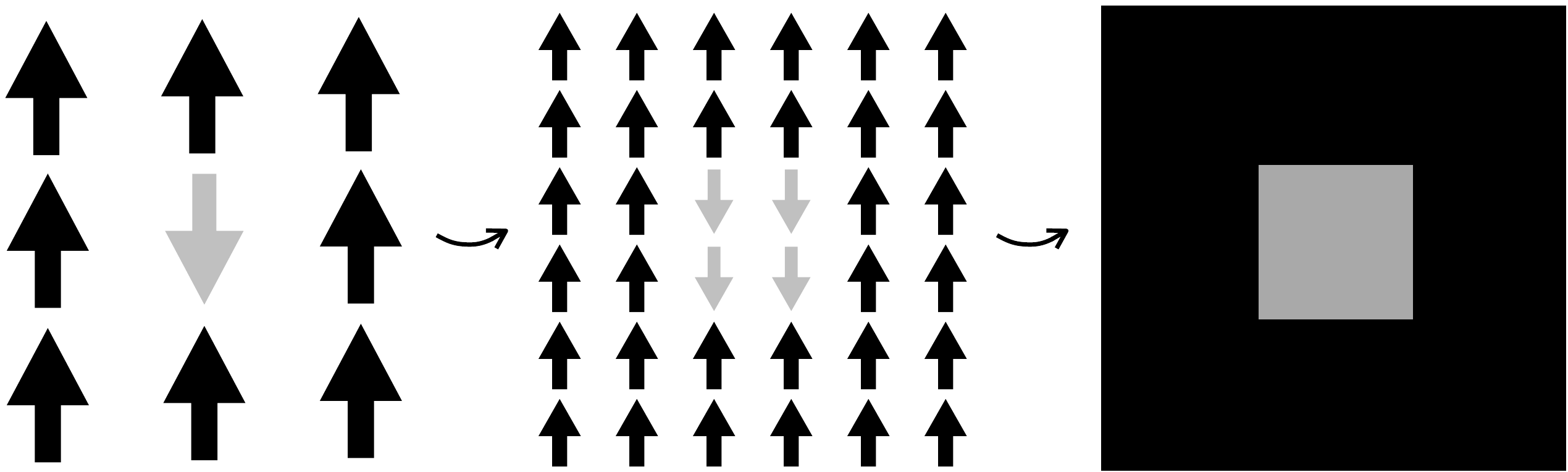}
\caption{Limit from discrete spin system to continuous system.}
\end{figure}

First, we introduce a spatial dimension along which magnetization may vary and take the coordinate $\xi\in[-1,1]$. A coupling kernel $G(\xi, \xi')$ determines the interaction between spatial coordinates $\xi$ and $\xi'$ (assumed to be periodic), and we retain $J$ as a constant coefficient, representing mean interaction energy for normalized $G(\xi,\xi')$. Next, we further generalize the traditional binary spins $s=\pm 1$ to allow continuous $s \in [-1,1]$. We then write
\begin{equation*}
  E = -\frac{1}{2}J \int_{-1}^1 \int_{-1}^1\!\! G(\xi,\xi')s(\xi)s(\xi')\,d\xi\,d\xi' - H \int_{-1}^1 \!\!s(\xi')\,d\xi'.
\end{equation*}

Computing the response to a differential change in $s$ at a single point in space, we find 
\begin{equation}
  \frac{\partial E}{\partial s} = -J \int_{-1}^1 G(\xi,\xi') s(\xi')\,d\xi' - H
  \label{dEdscts}
\end{equation}
(see Supplemental Material Section S1 for details on this derivation).

To describe the magnetization of the system, we use a ``mass action'' rate equation previously applied to sociophysics \cite{abrams2011}:
\begin{equation*}
  \frac{\partial x}{\partial t} = (1-x)P(x;u_x) - x P(1-x;1-u_x)~.
\end{equation*}
Here $x$ is the fraction belonging to one of two groups that partition the population and $u_x$ represents a relative bias towards this group. The function $P(x;u_x)$ gives the probability per unit time that an individual will switch into this group. We adapt this equation to ferromagnetism first by making the change of variable $x=(s+1)/2$ so that the magnetization $s\in[-1,1]$. We reinterpret the function $P$ to now represent the probability per unit time that a particle in the magnetic material will switch polarity, now depending on an applied magnetic field $H$ instead of the social bias $u_x$. 

In the discrete system, the probability that a particle will switch polarity is given by Glauber \cite{glauber1963} as inversely proportional to $1 + \exp\left(\frac{\Delta E}{kT}\right)$, where $\Delta E$ is the incremental change in energy caused by the spin flip, $k$ is Boltzmann's constant, and $T$ is temperature. Using the relationship $\Delta E\sim2\frac{\partial E}{\partial s}$ (each discrete spin flip $\Delta s$ has magnitude 2), we modify the Glauber rate to depend on differential changes in spin:
\begin{equation}
  P_G(s;H) \propto \left[ 1 + \exp\left( \frac{2}{kT}\frac{\partial E}{\partial s} \right) \right]^{-1}~.
  \label{fliprate}
\end{equation}

Using this expression for spin flip probability, our equation becomes 
\begin{equation}
  \frac{\partial s}{\partial t} = (1-s)P_G(s;H) - (1+s)P_G(-s;-H)~.
  \label{sODE}
\end{equation}
Note that time can be arbitrarily rescaled---a prefactor with units of inverse time has been set to one for convenience.

We may now write our full ODE system, combining the results of Eq.~\eqref{dEdscts}, \eqref{fliprate},  and \eqref{sODE}. After some simplification, we find the integrodifferential equation describing system magnetization $s(\xi,t)$ for a general coupling kernel:
\begin{equation}
  \frac{\partial s}{\partial t} = \tanh\!\left( \frac{J}{kT}\! \int_{-1}^1\!G(\xi,\xi')s(\xi',t)\,d\xi' + \frac{H}{kT} \right) - s~. 
  \label{sODE_full}
\end{equation}

\PRLsec{Homogeneous System}%
In the case of spatially homogeneous magnetization, where $s(\xi,t)=m(t)$, Eq.~\eqref{sODE_full} reduces to
\begin{equation}
  \dot{m}=\tanh\left[\frac{1}{kT}\left(Jm+H\right)\right] - m~.
  \label{homogeqn}
\end{equation}
The fixed points of this equation are equivalent to solutions of the mean-field Ising model \cite{domb1974}, but Eq.~\eqref{homogeqn} also allows for prediction of dynamics.  These dynamics are equivalent to the expected mean-field dynamics derived in \cite{suzuki1968} using a master equation approach. 

\begin{figure}
  \includegraphics[width=.75\linewidth]{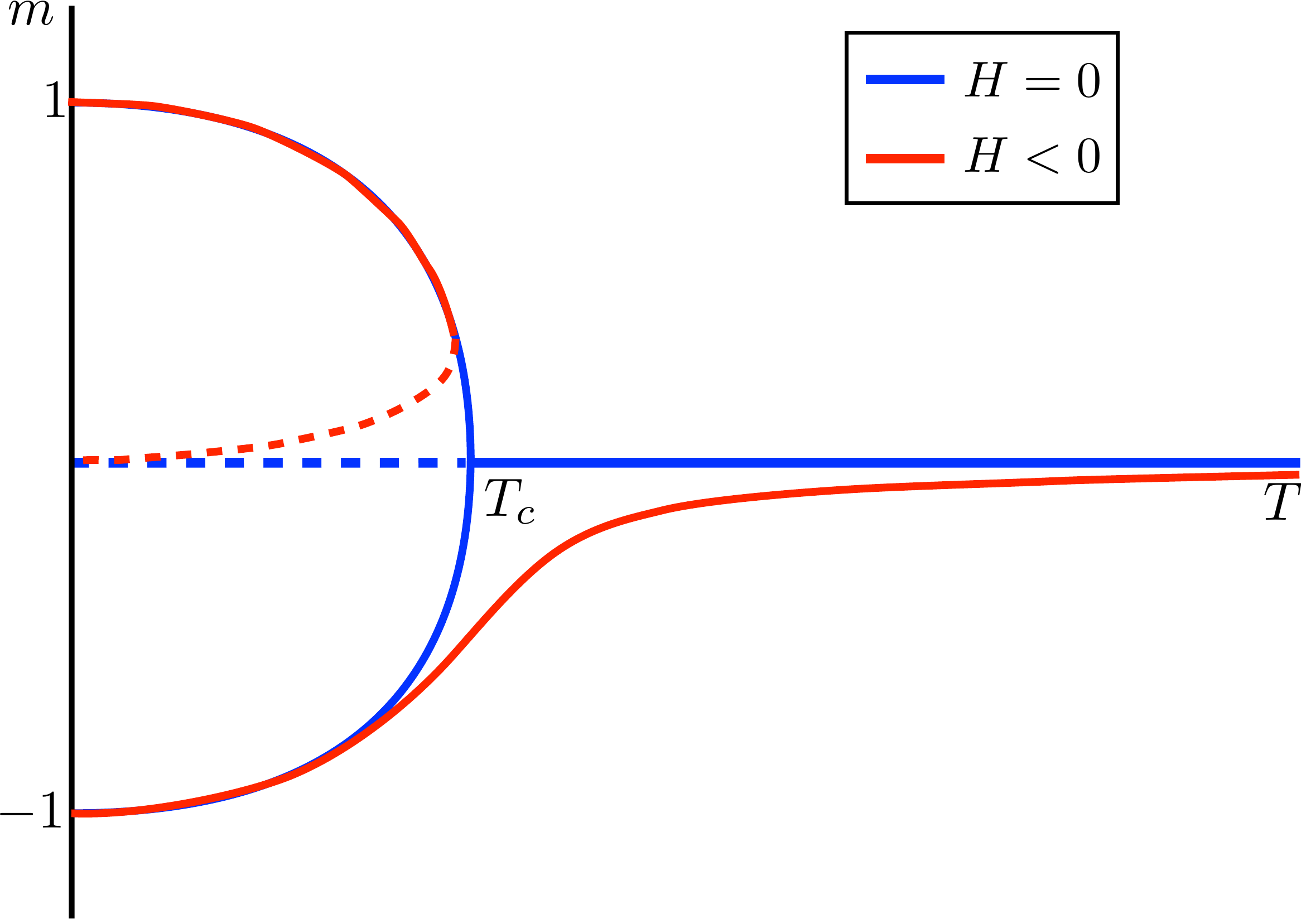}
  \caption{Steady-state magnetization versus temperature for homogeneous system, Eq.~\eqref{homogeqn}. For $H = 0$ there is a supercritical pitchfork bifurcation, while for $H \neq 0$, the bifurcation becomes imperfect. Solid lines indicate stable solution branches, dashed unstable.}
  \label{bifdiag}
\end{figure}

Analysis of Eq.~\eqref{homogeqn} reveals a supercritical pitchfork bifurcation, as is expected (see Fig.~\ref{bifdiag}). For zero temperature, the system must approach one of two stable fixed points $m^* = \pm 1$, while for temperatures above the bifurcation point $T_c = J/k$ (the Curie temperature), the sole fixed point is an unmagnetized (or ``thermalized'') state, $m^*=0$. This solution also exists for $T<T_c$, but is unstable. Including a nonzero external field gives an imperfect pitchfork bifurcation, where $m=0$ is no longer a solution for finite nonzero temperatures. These fixed points are given implicitly by solutions to $T=(J m^*+H)/(k\arctanh m^*)$.

\PRLsec{Piecewise-Constant System}%
The advantage of our formulation becomes clear when we relax the assumption of spatial homogeneity.  As an illustrative example, we consider a perturbation off all-to-all coupling towards local coupling, assuming piecewise-constant spins. One possible application of such a system is in modeling regions of a ferromagnet separated by grain boundaries---perhaps of relevance for magnetic memory applications.

We let $s(\xi,t)$ be piecewise constant and assume that intra-grain coupling is stronger than inter-grain coupling:
\begin{align}
  s(\xi,t)    &= 
    \begin{cases}
      m_1(t) &\text{if } \xi < 0\\
      m_2(t) &\text{if } \xi > 0
    \end{cases}\\
  G(\xi,\xi') &= \frac{1}{2}(1+\delta\,\text{sgn}\,\xi\,\text{sgn}\,\xi')~. \label{eq:piecewiseG}
\end{align} 
Then Eq.~\eqref{sODE_full} becomes
\begin{align*}
  \dot{m}_1&=\tanh\left\{\frac{J}{2kT}\big[m_1(1+\delta)+m_2(1-\delta)\big]+\frac{H}{kT}\right\}-m_1\\
\dot{m}_2&=\tanh\left\{\frac{J}{2kT}\big[m_2(1+\delta)+m_1(1-\delta)\big]+\frac{H}{kT}\right\}-m_2.
\end{align*}
The coupling strength $\delta$ acts as a perturbation parameter, where $\delta=0$ recovers Eq.~\eqref{homogeqn} for $\overline{m}=(m_1+m_2)/2$ and $\delta=1$ separates the system into independent halves that each evolve separately according to Eq.~\eqref{homogeqn}. Thus, intermediate values $\delta\in(0,1)$ are of interest.

\begin{figure}[htp]
  \centering
  \includegraphics[width=.9\linewidth]{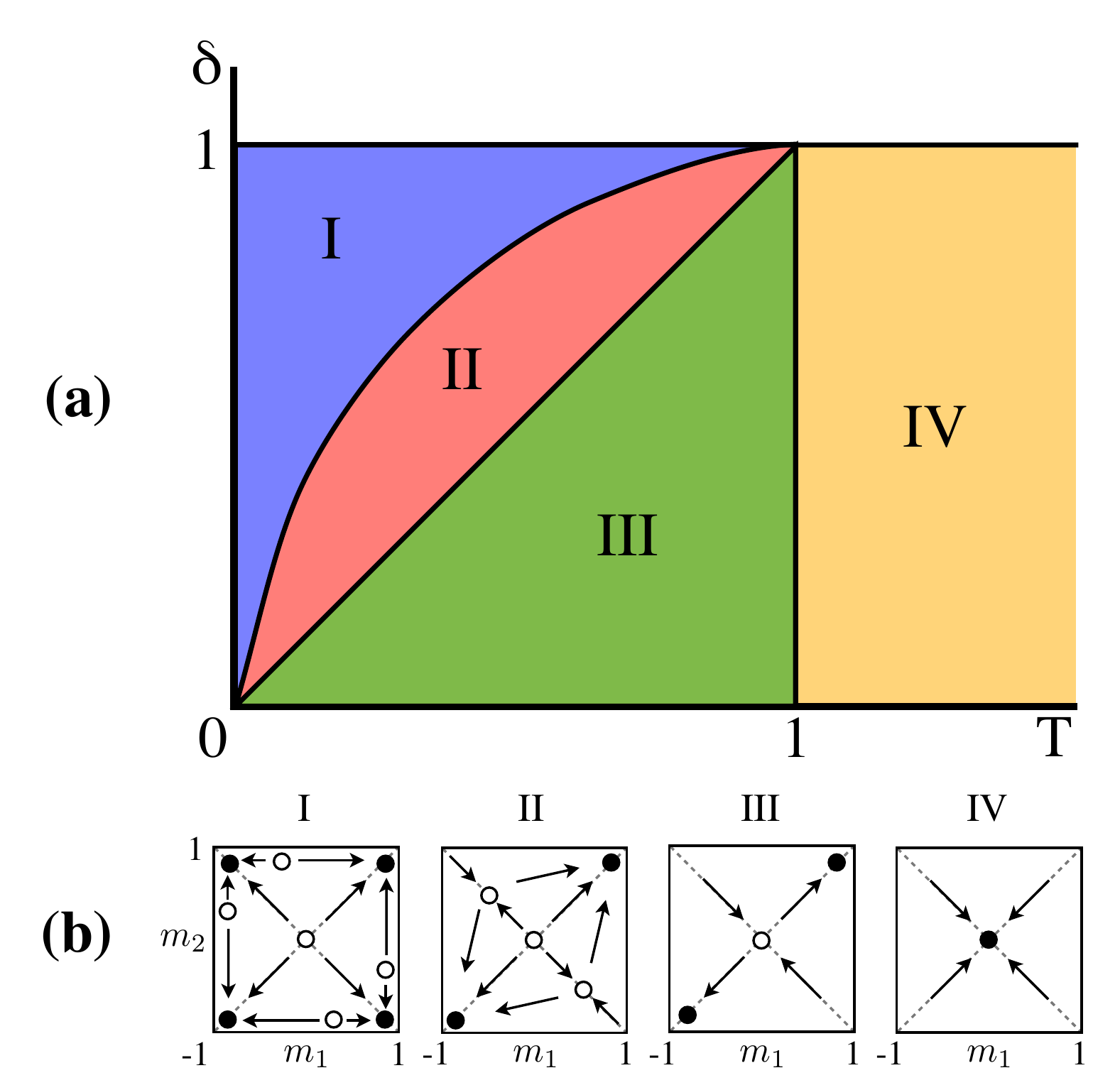}
  \caption{Fixed point locations and stability for piecewise-constant system. Shown in (a) are the regions of parameter space exhibiting qualitatively different behavior; the boundary between regions I and II is given approximately by $\delta=\frac{kT}{J} (1-\frac{kT}{J})^{-\frac{1}{2}}\arctanh (1-\frac{kT}{J})^{\frac{1}{2}}$. In (b), fixed point locations and stability are shown in phase space for each region; filled circles indicate stability, open circles instability. Arrows indicate direction of flow.
  \label{Tdeltaplane}}
\end{figure}

\PRLsec{Fixed Points for Piecewise System}%
For all temperatures and all $\delta$, fixed points exist on the invariant manifold $m_1=m_2$, where the two halves of the system are aligned and evolve according to Eq.~\eqref{homogeqn}.  Nontrivial fixed points may also be found on the invariant manifold $m_1=-m_2$: for $T<\delta$, besides $m_1^*=-m_2^*=0$, fixed points are also located at positions defined implicitly by $T=(J \delta m_1^*+H)/(k\arctanh m_1^*)$. Those points are stable for $\delta>\frac{kT}{J} (1-\frac{kT}{J})^{-\frac{1}{2}}\arctanh (1-\frac{kT}{J})^{\frac{1}{2}}$, otherwise they become saddle points that merge at $m_1=-m_2=0$ as $T \to \delta$.  See Figure~\ref{Tdeltaplane} for a visualization of how fixed point stability changes with $T$ and $\delta$.

The existence of stable fixed points on the $m_1=-m_2$ manifold is difficult to predict with the traditional approach to the Ising model, but has important implications: it indicates that a critical temperature exists below which two ferromagnetic grains will not equilibrate.

\PRLsec{Dynamics for Piecewise System}%
One benefit of modeling this system using a differential equation is that time-dependence is built in; we can evaluate the time scale of the approach to equilibrium by examining the system's eigenvalues, which  are especially straightforward to calculate in the piecewise constant system. For the fixed point at $m_1^*=m_2^*=0$ we find
\begin{align*}
  \lambda_1 &= -1+\frac{J\delta}{kT},\quad \mathbf{v}_1=(-1,1)\\
  \lambda_2 &= -1+\frac{J}{kT},      \quad \mathbf{v}_2=(1,1)~.
\end{align*}
Thus, along the antisymmetric manifold, the local time scale is $\tau_1 \propto T/(T-\delta T_c)$, and along the symmetric manifold, the local time scale is $\tau_2 \propto T/(T-T_c)$. These are relative thermalization times starting from either anti-aligned ($\tau_1$) or aligned ($\tau_2$) initial conditions.
 
Also of interest is the time it would take an unpolarized material to reach an equilibrium magnetization. In this case, it is straightforward to find the eigenvalues in terms of a general fixed point, either, $(m_1^*,m_2^*)=(m^*,m^*)$ or $(m_1^*,m_2^*)=(m^*,-m^*)$ .  We then find time scales $\tau\propto(1-c\frac{1-m^{*2}}{m^*}\text{arctanh}\,m^*)^{-1}$, where $c=1$ for flow perpendicular to the invariant manifolds,  $c=\delta$ for flow along the aligned manifold (meaning $\tau$ increases with $\delta$), and $c=1/\delta$ for flow along the anti-aligned manifold (meaning $\tau$ decreases with $\delta$). In each case, the flow towards equilibrium is slower when $m^*$ is closer to zero.

Note that this analysis gives us insight into the different dynamics observed in regions II and III of Figure~3. Despite sharing similar stable equilibria, the structural change in unstable fixed points affects the time scale for equilibration. Observation of new time scales in numerical simulation would be difficult to account for without a theory that includes dynamics.

\PRLsec{Von Mises Coupling}%
Although the simplicity of the piecewise constant system is appealing, it does not allow for smooth tuning between global (i.e. mean-field) and localized coupling. One coupling kernel that does allow such a continuous transition is the von Mises distribution
\begin{equation}
  G(\xi,\xi') = c e^{\kappa\cos\left[\pi (\xi-\xi')\right]}~,
  \label{eq:vonmisesG}
\end{equation}
with normalization constant $c=1/[2\, I_0(\kappa)]$. This distribution is similar to a Gaussian, but periodic. When the  parameter $\kappa$ is large, the coupling is local, while $\kappa=0$ recovers global coupling. 

\begin{figure}
  \centering
    \includegraphics[width=\linewidth]{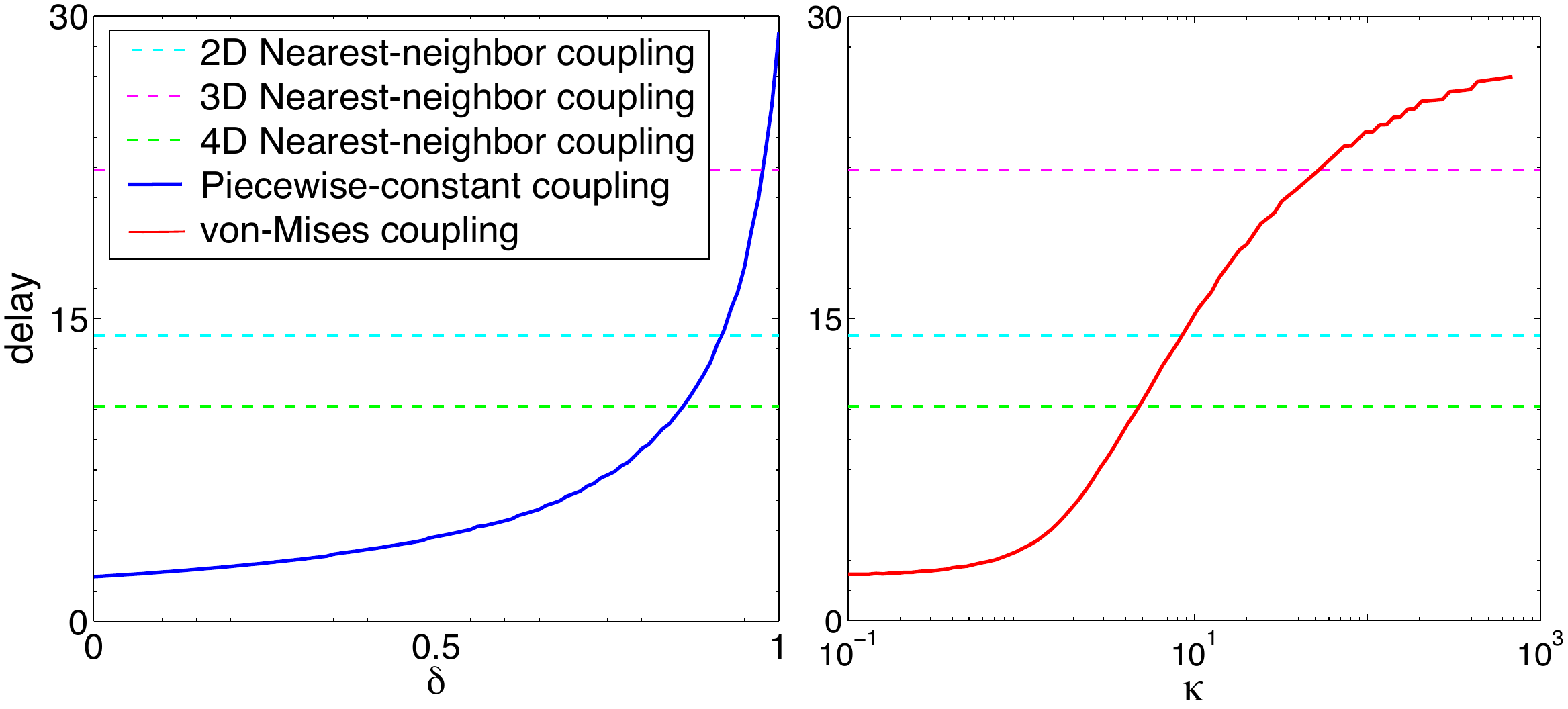}
  \caption{Equilibration time versus perturbation from mean-field coupling. Shown is the simulation time required for the system to evolve from 10\% to 90\% of the way between its initial condition and the thermalized (zero net magnetization) steady-state. Left panel: piecewise-constant coupling \eqref{eq:piecewiseG}; right panel: von Mises coupling \eqref{eq:vonmisesG}. Also shown, for reference, is the equilibration time for 2, 3, and 4-dimensional nearest-neighbor coupling (higher dimensions equilibrate faster).}
  \label{detdelay}
\end{figure}

We examined the equilibration time scale for this system, and found that the delay versus $\kappa$ curve is sigmoidal, behaving qualitatively like the piecewise constant system for $\kappa$ small. These results are shown in Figure~\ref{detdelay}.  Intriguingly, for both the piecewise constant and von Mises coupling kernels, the locality can be tuned to produce the same equilibration delay as that observed on a nearest-neighbor lattice of arbitrary dimension.

\PRLsec{Nearest-Neighbor Coupling}%
Conventionally, the Ising model couples nearest-neighbor spins arranged on a $d$-dimensional lattice. We can approximate this discrete coupling scheme in a continuous setting by defining our coupling kernel in terms of Dirac delta functions. For one-dimensional nearest-neighbor coupling (periodic), we use:
\begin{equation*}
  G(\xi,\xi') = \frac{1}{2} \left[ \delta(\xi-\xi'+\epsilon) + \delta(\xi-\xi'-\epsilon) \right],
\end{equation*}
where the prefactor is a normalization coefficient and $\epsilon \ll 1$ is analogous to grid spacing in the discrete system. For $H=0$, $T>T_c$ and $|s|$ small, Eq.~\eqref{sODE_full} then becomes
\begin{align*}
  \dot s 
    &= \tanh\left[ \frac{J}{2kT}\left( s(\xi+\epsilon)+s(\xi-\epsilon) \right) \right] - s\\
    &\approx \tanh\left[ \frac{J}{2kT}\left( 2s + \epsilon^2  s_{\xi\xi} \right) \right] - s\\
    & \approx \tanh\left[ \frac{Js}{kT} \right] - s + \frac{\epsilon^2 J}{2kT}s_{\xi\xi}~.
\end{align*}

In general, for $d$-dimensional nearest-neighbor coupling,
\begin{equation}
    \dot s \approx\tanh\left[ \frac{Js}{kT} \right] - s + \frac{\epsilon^2J}{2dkT} \nabla^2 s~.
  \label{diffusioneqn}
\end{equation}  
Note that the first two terms of the right-hand side are identical to those in Eq.~\eqref{homogeqn}, while the last term can be interpreted as introducing diffusion or smoothing to the system. Thus the nearest-neighbor approach may be approximated by combining uncoupled evolution towards the mean-field steady state and diffusion of magnetization across the system. Figure~\ref{diffusion} shows that this approximation works quite well for nearest-neighbor coupling (see Supplemental Material Section S3 for a discussion of the details of this approximation).

\begin{figure}
  \centering
  \includegraphics[width=.75\linewidth]{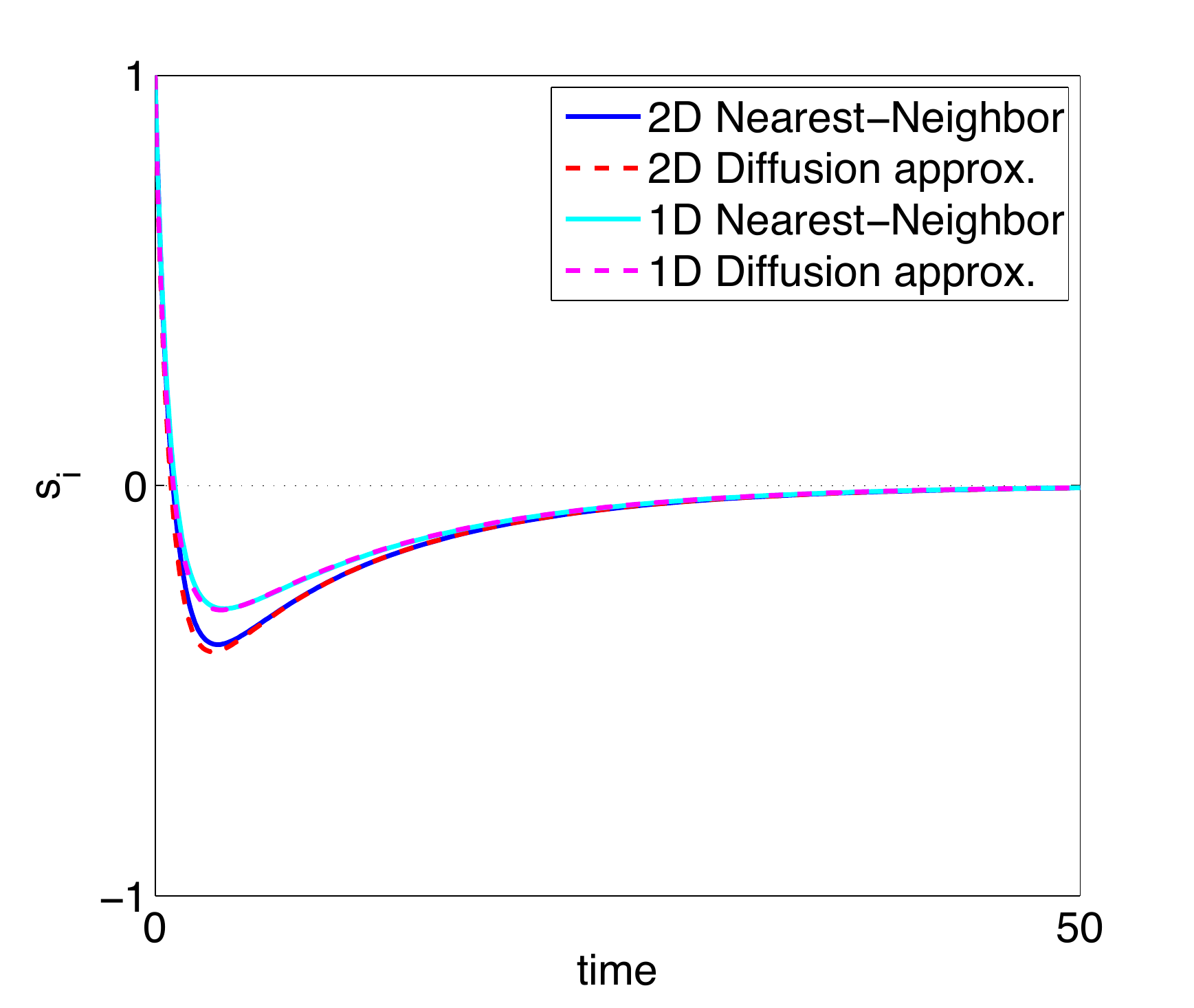}
  \caption{Perturbed spin versus time.  To test the approximation to the dynamics in Eq.~\eqref{diffusioneqn}, the system was initialized with a single spin $s_i = +1$ while the rest of the system was aligned at $-1$. The time evolution of this misaligned spin is shown for 1D and 2D nearest-neighbor systems together with the diffusive approximation. (The curves in 3D are similar and are omitted for clarity.)}
  \label{diffusion}
\end{figure}

\PRLsec{Material Properties}%
Another advantage of our model is that it allows for straightforward calculation of certain material properties. For example, the magnetic susceptibility $\chi=dm^*/dH$ is a measure of how magnetization is influenced by an externally applied field. We may calculate this directly by taking Eq.~\eqref{homogeqn} at steady-state (where $\dot m=0$ and $m(t)=m^*$)
and implicitly differentiating with respect to $H$, resulting in the following expression: 
\begin{equation*}
  \chi=\left\{k T \cosh^2 \left[ \frac{1}{kT} \left(Jm^*+H \right) \right] - J\right\}^{-1}~.
\end{equation*}
In the limit where $m^*\to0$ and $H\to0$, we find $\chi\propto(T-J/k)^{-1}$, matching the known result of $\chi\propto(T-T_c)^{-\gamma}$ with the mean-field value $\gamma=1$ \cite{kadanoff1967}.

For 2D and 3D nearest-neighbor systems, we know from theoretical, numerical, and experimental work that $\gamma>1$ \cite{kadanoff1967}.  A drawback to our approach is that the integrodifferential equation \eqref{sODE_full} cannot give $\gamma \neq 1$ for any smooth coupling kernel.  However, an alternative method of gaining insight into this critical exponent may be possible (see Supplemental Material Section S2).

\PRLsec{Conclusions}%
We have taken an unusual approach in our generalization of the Lenz-Ising system via continuum limits in space and spin.  Our model discards stochastic fluctuations and therefore precludes the use of many tools of statistical mechanics, but it allows the introduction of a different set of tools typically associated with nonlinear deterministic dynamical systems.   

An advantage to our approach is that, once formulated, our model can be analyzed with straightforward calculations that yield fixed points, stability, dynamics, and material properties for arbitrary coupling configurations.  The model is amenable to perturbative analysis, allowing for analytical predictions and generating some insight even in cases where exact solution is not possible.  
For example, with piecewise-constant coupling our model predicts disjoint a equilibrium below a threshold temperature, something that has not been found with the traditional mean-field treatment in statistical mechanics.
Because the model is deterministic, a large class of well-explored methods work well in numerical solution.

A disadvantage of this model is that smooth coupling kernels cannot exactly reproduce the behavior of discrete coupling arrangements such as nearest-neighbor.  However, it remains unclear whether a smooth approximation to nearest-neighbor coupling may in fact allow for approximation of more complete results such as non-mean-field critical exponents (see Supplemental Material Section S2 for more discussion of this point). Regardless, we believe that this approach to Ising model systems may provide new insight into a long-standing problem.

\bibliographystyle{apsrev4-1}

%

\end{document}


\title{Supplemental Material:\\A Continuous Generalization of the Ising Model}

\author{Haley A. Yaple}
\email[email: ]{haleyyaple@u.northwestern.edu}
\affiliation{Department of Engineering Sciences and Applied Mathematics, Northwestern University, Evanston, IL 60208, USA}

\author{Daniel M. Abrams}
\affiliation{Department of Engineering Sciences and Applied Mathematics, Northwestern University, Evanston, IL 60208, USA}
\affiliation{Northwestern Institute on Complex Systems, Northwestern University, Evanston, IL 60208, USA}
\date{\today}

\maketitle

\section{Continuum Limit of Change in Energy}

\subsection{Derivation in continuous setting}
 In the continuum limit, it is natural to assume that Eq.~(1) becomes 
\begin{equation*}
E=-\frac{1}{2}J\int_{-1}^1\int_{-1}^1G(\xi,\xi')s(\xi)s(\xi')\,d\xi\,d\xi'-H\int_{-1}^1 s(\xi')\,d\xi'~,
\end{equation*}
where $G(\xi,\xi')$ gives the coupling between two points $\xi,\,\xi'\in[-1,1]$ and $s(\xi)$ is the value of the spin located at $\xi$. 
To find $\Delta E$, we consider changing the spin at just one point $\xi=a$ by letting $s(\xi)\to s(\xi)+\eta\delta(\xi-a)$. Then the new energy of the system is
\begin{align*}
E_\eta&=-\frac{1}{2}J\int_{-1}^1\int_{-1}^1G(\xi,\xi')\left[s(\xi)+\eta\delta(\xi-a)\right]\left[s(\xi')+\eta\delta(\xi'-a)\right]\,d\xi\,d\xi'-H\int_{-1}^1 \left[s(\xi')+\eta\delta(\xi'-a)\right]\,d\xi' \\
&= E-\frac{1}{2}J\eta\int_{-1}^1 G(\xi,a)s(\xi)\,d\xi-\frac{1}{2}J\eta\int_{-1}^1 G(a,\xi')s(\xi')\,d\xi' -\frac{1}{2}J\eta^2G(a,a)-H\eta\\
&=E-J\eta\int_{-1}^1 G(a,\xi')s(\xi')\,d\xi'-H\eta\\
\Delta E &= E_\eta-E =  -J\eta\int_{-1}^1 G(\xi,\xi')s(\xi')\,d\xi'-H\eta
\end{align*}

This gives us an expression for the change in energy in response to a change in spin of magnitude $\eta$. We assume in the limit $\eta\to0$ that $\frac{\Delta E}{\Delta s}\to \frac{\partial E}{\partial s}$, where $\eta\approx \Delta s$, and thus we find:
\begin{equation*}
\frac{\partial E}{\partial s}=-J\int_{-1}^1G(\xi,\xi')s(\xi')\,d\xi'-H~.
\end{equation*}

\subsection{Derivation in discrete setting}
In this section, instead of assuming the continuum limit naturally converts sums into integrals, we explicitly take this limit. We do this by beginning with the traditional discrete Ising model with $N$ spins, and replacing each spin with a cluster of size $M$. Then the effective spin at position $i$ is given by the average of the spins in cluster $i$:
\begin{equation}
m_i=\frac{1}{M}\sum_{k=1}^M s_{ik}. \quad\text{(here $s_{ik}$ is the $k$th spin in cluster $i$)}\label{effectivespin}
\end{equation}

Now compute the total system energy, replacing the sums over all spins in Eq.~(1) by sums over all clusters. We replace $J_{ij}$, the interaction energy between spins, by $G_{ij}$, the interaction energy between clusters. Because each interaction between clusters represents $\mathcal{O}(M^2)$ interactions between spins, we let $J_{ij}M^2=G_{ij}$. 
\begin{align*}
E&=-\frac{1}{2}\sum_{i=1}^N\sum_{j=1}^N G_{ij}m_im_j-HM\sum_{i=1}^N m_i
\end{align*}

Consider the effect of flipping one spin, taken without loss of generality to be within cluster $N$. Denote the effective cluster spin by $m_N$ before the flip and $m_N^*$ after the flip.
\begin{align*}
\Delta E&=E_\text{(after flip)}-E_\text{(before flip)}\\
&=\left[-\frac{1}{2}\sum_{i=1}^{N-1}\sum_{j=1}^{N-1} G_{ij}m_im_j -HM\sum_{i=1}^{N-1} m_i -\frac{1}{2}\left(\sum_{i=1}^{N-1}G_{iN}m_im_N^*+\sum_{j=1}^{N-1}G_{Nj}m_N^*m_j+G_{NN}(m_N^*)^2\right)-HMm_N^*  \right]\\&\quad-\left[-\frac{1}{2}\sum_{i=1}^{N-1}\sum_{j=1}^{N-1} G_{ij}m_im_j -HM\sum_{i=1}^{N-1} m_i -\frac{1}{2}\left(\sum_{i=1}^{N-1}G_{iN}m_im_N+\sum_{j=1}^{N-1}G_{Nj}m_Nm_j+G_{NN}m_N^2\right)-HMm_N \right]\\
&=-\sum_{i=1}^{N-1}G_{iN}m_i(m_N-m_N^*)-\frac{1}{2}G_{NN}(m_N^2-(m_N^*)^2)-HM(m_N-m_N^*)\\
&=(m_N-m_N^*)\left(-\sum_{i=1}^{N-1}G_{iN}m_i-\frac{1}{2}G_{NN}(m_N+m_N^*)-HM\right)
\end{align*}

Now consider the effect of flipping one spin on the effective spin of the cluster $m_N$, as calculated by Eq.~\eqref{effectivespin}. If the spin was negative, $m_N$ is increased by $2/M$, while if the spin was positive, $m_N$ is decreased by $2/M$. In general, $m_N-m_N^*=2s_k/M$, where $s_k$ is the original value of the spin that is to be flipped. Thus
\begin{align*}
\Delta E&=\frac{2s_k}{M}\left(\!-\sum_{i=1}^{N-1}G_{iN}m_i-G_{NN}\left(m_N-\frac{1}{M}\right)-HM\!\right)~.
\end{align*}
Now, in general we assume that the system has no self-coupling, i.e. $G_{NN}=0$. Then we have
\begin{align*}
\Delta E&=2s_k\left(-\sum_{i=1}^{N-1}G_{iN}m_i\frac{1}{M}-H\right)~.
\end{align*}

If we consider the first term to be a Riemann sum of the form $\sum_{i=1}^Nf(x_i)\Delta x$, we can see how our change in energy may be written as an integral. First, we require that the cluster size $M$ and the number of clusters $N$ approach infinity at the same rate. This is reasonable, as otherwise we would have, in the limit, either an infinite number of clusters of negligible size, or a negligible number of infinite-size clusters, neither of which is desirable. If we assume $M=cN$, the coefficient $c$ may be used to rescale temperature in our full system. Furthermore, we assume that in this limit $\frac{\Delta E}{\Delta s}\to \frac{\partial E}{\partial s}$, where $\Delta s$ is one change in spin, or $2s_k$. We rename the coupling interaction coefficient as $G_{ij}=G(x_i,x_j)$, which in the limit becomes a continuous function of two spatial variables $G(\xi,\xi')$, and similarly rename the effective spin $m_i$ as $s(\xi)$. Thus we have
\begin{align*}
\frac{\partial E}{\partial s}&=-\int_{-1}^1 G(\xi,\xi')s(\xi')d\xi'-H~.
\end{align*}

\section{Susceptibility}
Although continuous coupling kernels yield mean-field results for critical exponents in our model, clearly the dynamics do change as the locality of the coupling changes, as evidenced by Figure~4. It may be possible to infer properties of the stochastic Ising model from the dynamics of our deterministic generalization.

In the typical statistical mechanical approach, for example, susceptibility may be calculated as $\chi = \var(m^*) / T$, where the variance is over an ensemble of realizations. Clearly, as our model is deterministic, we cannot have such variance. We instead look to the equilibration time as a proxy, specifically the time scale on which the system approaches a homogeneous spin distribution. That time scale is related to the equilibration time in the original Ising system, which is in turn related to the susceptibility $\chi$.  This may allow us to extract a relationship between the deterministic equilibration time and the susceptibility of the equivalent stochastic system. Preliminary numerical experiments do suggest that this approach holds promise.


\section{Dimensionality}

The diffusion constant that appears in Equation~(10) of the main text is $D=\epsilon^2 J / (2dkT)$.  It might seem that this diffusivity decreases with the dimension $d$, but in fact the behavior is exactly the opposite.

In a typical implementation of the traditional Ising model, the net coupling increases with the dimensionality: that is, interaction energy $J_{ij}$ remains constant as the number of nearest neighbors increases, so net interaction energy for a polarized system is proportional to $d$.  We can capture this effect in our generalization by setting $J \propto d$. 

Our parameter $\epsilon$ should also vary with the dimensionality of the system.  For two Ising model systems of identical physical size and identical numbers of $N$ spins, but different dimensions, the ``grid spacing'' $\epsilon$ must vary as $\epsilon \propto N^{-1/d}$.

Thus we expect the effective diffusion constant for equivalent Ising systems to \textit{increase} with dimension as $D \propto \exp{(-2/d)}$.  This is reasonable, given that we expect information to diffusive more effectively on a highly interconnected network than on a sparse network.